\documentstyle[eqsecnum,twocolumn,aps]{revtex}
\begin{document}
\draft
\title {Nonequilibrium Phase Transition in the Kinetic Ising model:\\
 Critical Slowing Down and Specific-heat Singularity}
\author {Muktish Acharyya$^*$}
\address {Department of Physics\\
Indian Institute of Science, Bangalore-560012, India\\
 and\\
 Condensed Matter Theory Unit\\
 Jawaharlal Nehru Centre for Advanced Scientific Research\\
 Jakkur, Bangalore-560064, India}
\maketitle
\narrowtext
\begin{abstract}
The nonequilibrium dynamic phase transition,
in the kinetic Ising model 
in presence of an oscillating magnetic field,
has been studied both by Monte Carlo
simulation and by solving numerically the mean field dynamic equation
of motion
for the average magnetisation. In both the cases, the 
Debye 'relaxation' 
behaviour of the dynamic order parameter has
been observed and the 'relaxation
time' is found to diverge near the dynamic transition point.
The Debye relaxation
of the dynamic order parameter and the
power law divergence  
of the relaxation time have been
obtained from a very
approximate solution of the mean field dynamic equation.
The temperature variation of 
appropiately defined 'specific-heat' 
is studied by Monte Carlo simulation near the transition point. The 
specific-heat
has been observed to diverge near the dynamic transition point.
\end{abstract}


\section{Introduction}

The dynamic response of the Ising system in presence of an oscillating 
magnetic field has been studied
extensively [1-3] in the last few years. The dynamic hysteresis 
[1-3] and the nonequilibruim
dynamic phase transition [4-6] are two important aspects of
the dynamic response of the kinetic Ising model in presence of an
oscillating magnetic field. This kind of phase transition 
in the Ising model
was first studied by Tome and Oliviera [4]. They solved the mean field (MF)
dynamic equation of motion (for the average magnetisation) of the kinetic
Ising model in presence of a sinusoidally oscillating magnetic field.
They have defined the order parameter as the time averaged
magnetisation over a full cycle of the 
oscillating field and showed that depending upon
the value of the field and the temperature, the order parameter takes
nonzero value from a zero value. 
Precisely, in the field amplitude and temperature plane there
exists a phase boundary separating dynamic ordered 
(nonzero value of order parameter) and disordered (order 
parameter vanishes) phase. They [4] have also observed 
and located a tricritical point (TCP),
(separating the nature (discontinuous/continuous) of the transition)
on the phase boundary line (see Fig.1). However, such a mean field transition
is not truely dynamic in origin and exists even in the static (or zero
frequency) limit. This is because, if the field amplitude is less than the
coercive field (at temperature less than the transition temperature
without any field), then the response magnetisation varies periodically
but asymmetrically even in the zero frequency limit; the system remains locked
to one well of the free energy and cannot go to the other one, in the absence
of fluctuation.

Lo and Pelcovits [5] first attempted to study the dynamic 
nature of this phase transition
in the kinetic Ising model by Monte Carlo (MC) simulation. 
Here, the transition disappears in the 
zero frequency limit; due to the fluctuations, the magnetisation flips to the
direction of the magnetic field and the dynamic order parameter (time
averaged magnetisation) vanishes.
However, Lo and Pelcovits [5] have 
not reported any precise phase boundary.
Acharyya and Chakrabarti [6] studied the nonequilibrium dynamic phase
transition in the kinetic Ising model 
in presence of oscillating magnetic field by 
extensive MC simulation. 
They [6] have also identified 
that this dynamic phase transition is associated with the 
breaking of the symmetry of
the dynamic hysteresis ($m-h$) loop. 
In the dynamically disordered (value of order parameter vanishes)
phase the corresponding hysteresis loop is
symmetric, and loses its symmetry in the ordered phase (giving
nonzero value of dynamic order parameter).
They [6] also studied the temperature variation of the ac susceptibility
components near the dynamic transition point.  
The major observation was
that the imaginary (real) part of the ac susceptibility gives a
peak (dip) near the dynamic transition point (where the dynamic
order parameter vanishes). The conclusions were: (i) this is a distinct
signal of phase transition and (ii) this is an indication of the
thermodynamic nature of the phase transition.

It is worth mentioning here that the statistical distribution of dynamic
order parameter has been studied by Sides et al [7]. The nature of the
distribution changes near the dynamic transition point. They have
also observed 
[7] that the fluctuation of the hysteresis loop area becomes considerably
large near the dynamic transition point.

In the equilibrium critical phenomena, both the length
and time scales diverge at criticality. 
This gives rise to the singularities in various thermodynamic quantities,
like the specific heat and the relaxation time.
Can one expect similar kind of features in the case of this nonequilibrium
dynamic phase transition problem ?
To be specific, (i) Is there any 'relaxation time', in this nonequlibrium
problem which diverges at the dynamic transition point ? (ii) Can there be any
appropiately defined 'specific-heat', which will
show singular behaviour at the
transition point ?

The 
main motivation of this paper is to get some answers of these 
questions at least numerically.
The nonequilibrium dynamic phase transition in the kinetic Ising model in
presence of an oscillating magnetic field has been studied by MC 
simulation. Also the MF dynamic equation has been solved numerically, 
to compare the results.
The 'relaxation' behaviour 
(defined in the following section)
of the dynamic order parameter [4] and the behaviour
of 'specific-heat' (defined in section III)
near the dynamic transition point are
studied by MC simulation. 
It may be mentioned here that the preliminary results of 'specific-heat'
singularity near the dynamic transition point were reported briefly in ref [8].
More detailed results are reported here.
The 'relaxation' behaviour
has also been studied here by solving numerically the mean field (MF)
dynamic equation of
motion of the kinetic Ising model in
presence of an oscillating magnetic field. 
The MF equation has also been solved 
exactly in the linearised limit and studied the 
'relaxation' behaviour of the dynamic order parameter, near the dynamic 
transition point.
The paper is organised as follows:
In section II the 'relaxation' behaviour of the order parameter 
near the dynamic transition point has been
studied both by Monte Carlo simulation and by solving mean field dynamic
equation of motion of the kinetic Ising model. 
In section III the temperature variation of the 
 'specific-heat' 
has been studied 
near the transition point only by Monte Carlo simulation. A brief summary
of all the results is given in section IV.

\section{'Relaxation' behaviour of the dynamic order parameter}
\subsection{Monte Carlo Study}
\subsubsection{The Model and Simulation}

A ferromagnetically interacting (nearest neighbour) Ising model in presence
of a time varying magnetic field can be represented by the Hamiltonian

\begin{equation}
H = -\sum_{<ij>} J_{ij} s_i^z s_j^z - h(t) \sum_i s_i^z
\label{hm} 
\end{equation}
Here, $s_i^z (=\pm 1)$ is Ising spin variable, $J_{ij}$ is the interaction
strength and $h(t) = h_0 {\rm cos}(\omega t)$  
represents the oscillating magnetic field, where
$h_0$ and $\omega$ are the amplitude and the frequency 
respectively of the oscillating field. The system
is in contact with an isothermal heat bath at temperature $T$. For simplicity
all $J_{ij}$ are taken equal to 
unity and periodic boundary condition is chosen.

A square lattice of linear size $L (=100)$ has been considered.
At any finite
temperature $T$ and for a fixed frequency ($\omega$) 
and amplitude ($h_0$) of the
field, the 
dynamics of this system has been studied here by Monte Carlo
simulation using Glauber single spin-flip dynamics
with the Metropolis rate of spin flip. Each lattice site is
updated here sequentially and one such full scan over the lattice is
defined as the time unit (Monte Carlo step or MCS)
here. The initial configuration has been chosen
such that the all spins are directed upward. The instanteneous magnetisation
(per site),
 $m(t) = (1/L^2) \sum_i s_i^z$ has been calculated. From the instanteneous
magnetisation, the dynamic order parameter $Q = {\omega \over {2\pi}}
\oint m(t) dt$ (time averaged magnetisation over a full cycle of the
oscillating field) is calculated. 

\subsubsection{Results}

Fig.1 shows the schematic diagram of the dynamic phase boundary
in the field amplitude ($h_0$) and temperature ($T$) plane. 
For small values of $h_0$ and $T$ the dynamic order parameter $Q$ is
nonzero and the corresponding dynamic hysteresis loop ($m-h$ loop) is
asymmetric and for larger values of $h_0$ and $T$ the dynamic order
parameter $Q$ vanishes, correspondingly, the $m-h$ loop becomes 
symmetric
(inset of Fig.1). 
The dynamic transition temperature ($T_d$) is a function of field
amplitude ($h_0$).
The transition across the dotted line (in Fig.1) is
discontinuous and that across the solid line is continuous. For very small
values of $h_0$ the nature of the dynamic transition is continuous. 
In this
paper, all studies are done in the region where $Q$ undergoes always
a continuous transition.

It has been observed carefully that the dynamic order parameter $Q$
does not acquire the stable value within the first cycle of the oscillating
field. It takes several cycles (of the oscillating field) to get
stabilised i.e., it shows 'relaxation' behaviour.
Starting from the initial (all spins are up) configuration, the $Q$ 
has been calculated for various number (say $n$-th) of cycles of the oscillating
magnetic field and plotted (inset of Fig.2) against the number of cycles ($n$). 
Each value of $Q$ shown here has been obtained by averaging over 100 random
Monte Carlo samples. 
Inset of
Fig. 2 shows a typical 'relaxation' 
behaviour of the dynamic order parameter $Q$. 
This has been plotted for fixed values
of $\omega = 2\pi \times 0.04$, 
$h_0$= 1.0 and  $T$= 1.5. It shows that the dynamic order
parameter $Q$ is relaxing as the time (number of cycles) goes on. 
The best fit curves shows that the 'relaxation' is exponential type.
So, one can write $Q \sim Q_0 {\rm exp}(-n/\Gamma)$, where $\Gamma$ is the
'relaxation' time which provides the 'time scale' for this nonequilibrium
problem.
The physical interpretation of $\Gamma$ is,
the number of cycles required, so that $Q$ becomes $1/e$ times of
its initial value (value of $Q$ at starting cycle). 
From the exponential
fitting, the 'relaxation' time ($\Gamma$) has been meseared.
The temperature ($T$) variation, for fixed
values of $\omega$ and $h_0$, of this 'relaxation' time 
$\Gamma$ has been studied
(in the disordered region of dynamic transition)
and displayed in Fig.2. 
The temperature ($T$) variation of $\Gamma$ are shown (Fig.2) for
two different values of $h_0 (= 0.5 {\rm~~and~~} 1.0)$ and for a fixed
value of $\omega = 2\pi \times 0.04$ here.
From the figure (Fig.2) it is clear that the relaxation time $\Gamma$ 
diverges near the dynamic transition point (where $Q$ vanishes) in the both
cases ($h_0$ = 0.5 and 1.0).

\subsection{Mean field study}
\subsubsection{Mean field equation of motion and numerical solution}

Although as mentioned earlier, the mean field system does not undergo
a true dynamic transition (as the transition exists even in the static limit),
the mean field case has been considered here as a pathological one.

The time evolution of the average magnetisation (under mean field 
approximation) in presence of an oscillating magnetic field can be
described by the equation

\begin{equation}
\tau {dm \over dt} = - m + {\rm tanh} 
\left({{m(t)+h(t)} \over T}\right),
\label{mf}
\end{equation}
where $m(t)$ is the instanteneous 
magnetisation, $h(t) = h_0 cos (\omega t)$ is
a sinusoidally oscillating magnetic field, 
$T$ is the temperature and $\tau$
is a constant.

This equation has been solved by fourth order Runge-Kutta method taking
$\tau = 2\pi \times 0.01$ and $dt$=0.01. 
The initial boundary condition is $m(0)=1.0$. From the numerical solution
for the instanteneous magnetisation $m(t)$, the dynamic order parameter
$Q (= {{\omega} \over {2\pi}}\oint m(t) dt)$ has been calculated.

\subsubsection{Results}

The inset of Fig.3 shows a typical
'relaxation' of the dynamic order parameter $Q$ for 
$\omega = 2\pi \times 0.02$, $h_0$ = 0.4
and $T$ = 0.765.
Here, also the exponential type of relaxation is observed and
the 'relaxation' time has been measured in the same way, discussed 
earlier (in the MC case).
Fig 3 shows the temperature variation of the 'relaxation' time
$\Gamma$ for $\omega = 2\pi \times 0.02$
and two different values of $h_0$ (= 0.3 and 0.4).
Here also from the figure (Fig.3) it is clear that the typical time scale 
or the 'relaxation' time $\Gamma$ for this nonequilibrium problem, 
diverges, for both the cases ($h_0$ = 0.3 and 0.4),
near the dynamic transition point (where $Q$ vanishes).

\subsubsection {An approximate solution of MF equation}

In the limit of $h_0 \to 0$ and $T > 1$, the equation (\ref{mf}) can be
linearised (i.e., linearising tanh term) as 
\begin{eqnarray*}
\tau {dm \over dt} = -\epsilon m + {{h_0 {\rm cos}(\omega t)} \over T},
\end{eqnarray*}
\noindent where $\epsilon = 1 - 1/T$.
The solution of the above equation is
\begin{eqnarray*}
m(t) = {\rm exp}(-\epsilon t/\tau) + m_0 {\rm cos}(\omega t -\phi),
\end{eqnarray*}
\noindent where $m_0$ and $\phi$ are two constants.
The value of the dynamic order parameter $Q$ at $n-$th cycle of the
oscillating field is
\begin{eqnarray*}
Q = {{\omega} \over {2\pi}} \oint m(t) dt = {{\omega} \over {2\pi}}
\int_{t_{n-1}}^{t_{n}} m(t) dt,
\end{eqnarray*}
\noindent where $t_n = 2\pi n/\omega$.
The value of $Q$, at the $n$-th cycle, can be written as
\begin{eqnarray*}
Q = Q_0 {\rm exp}(-{{2\pi n \epsilon} \over {\tau \omega}}) 
= Q_0 {\rm exp} (-n/\Gamma)
\end{eqnarray*}
\noindent $Q_0$ is a constant independent of $n$. The above form shows that
$Q$ relaxes exponentially with the number of cycles ($n$) of the oscillating
field. The 'relaxation' time $\Gamma$ is equal to ${{\tau \omega} \over
{2 \pi}} \epsilon^{-1}$. 
It should be noted here that the dynamic transition occurs at $T$ = 1 in the
limit $h_0 \to 0$ [4]. So, for $h_0 \to 0$  near the dynamic transition point 
(where the linearisation holds good) the behaviour of
relaxation time is
\begin{eqnarray*}
\Gamma \sim \epsilon^{-1} \sim \left(T - T_d(h_0 \to 0)\right)^{-1}
\end{eqnarray*}
\noindent which shows the power law (exponent is unity) divergence 
of the 'relaxation' time at
the dynamic transition point.

\section{Behaviour of 'specific-heat' near the transition point}

 The time averaged (over a full cycle) cooperative energy of the system
 may be defined as

$$(a) ~~~~E_{coop} = -(\omega/{2 {\pi} L^2}) \oint 
\left(\sum_{<ij>} s_i^z s_j^z \right) dt,$$

\noindent and the time averaged (over a full cycle) total 
 energy (including both cooperative and field part) of the system
 can be written as

$$(b) ~~~~E_{tot} = -(\omega/{2 {\pi} L^2})\oint 
\left(\sum_{<ij>} s_i^z s_j^z 
+ h(t) \sum_i s_i^z \right)dt.$$

\noindent The temperature variations 
of $E_{tot}$ and $E_{coop}$ have been studied.
The specific heat's are defined as $C_{tot} = dE_{tot}/dT$
and $C_{coop} = dE_{coop}/dT$.
The temperature variations of the 'specific-heat's have also been studied
and found prominent divergent behaviour near the dynamic transition point
(where $Q$ vanishes).

Here, again a square lattice of linear size $L$ (=100) has been
considered. Both $E_{coop}$ and $E_{tot}$ are calculated using MC simulation.
Each data point has been obtained by averaging over 100 different MC
samples.

\subsection{Results}

The temperature derivatives of $E_{coop}$ and $E_{tot}$ can be defined as the
'specific-heat's for this nonequilibrium problem.
The temperature variations of $Q$ , 
$C_{coop}(= dE_{coop}/dT$) and 
$C_{tot} (= dE_{tot}/dT$) have been studied. 
The values of $h_0$ (=0.4 and 0.8)
are chosen here in such a way that $Q$ always undergoes
a continuous transition.
The temperature variations of $Q$, 
$C_{coop}$  has been shown in Fig. 4. Inset shows the variation
of total cooperative energy $E_{coop}$ (per spin) with temperature ($T$).
In this case the frequency ($\omega$) of the field is kept fixed 
($\omega$ = 0.0628).  Fig. 5 shows the temperature variation (for the
same values of $\omega$, $h_0$ and $T$) of $C_{tot}$ and the inset shows
the temperature variation of the total (cooperative + field) energy
(per spin).
From the figure it is clear that the appropiately defined 'specific-heat' s
$C_{coop}$ and $C_{tot}$  
 diverge near the dynamic phase transition point.

\section{Summary}

The nonequlibrium dynamic phase transition, in the kinetic Ising model
in presence of oscillating magnetic field, is studied both by Monte
Carlo simulation and by solving the meanfield dynamic equation of motion.

Acharyya and Chakrabarti [6] observed that the complex susceptibility
components have peaks (or dips) at the dynamic transition point. Sides et
al [7] observed that the fluctuation in the hysteresis loop area grows
(seems to diverge) near the dynamic transition point.

In this study it is observed that the 'relaxation time' and the appropiately
defined 'specific-heat' diverge near the dynamic transition
point. All the results are obtained here numerically. No attempts were made
to extract any exponent values from the numerical studies.

It should be mentioned that recent experiments [9] on ultrathin
ferromagnetic Fe/Au(001) films have 
been performed to investigate the frequency dependence
of hysteresis loop areas.
Recently, attempts have been made [10] to measure
the dynamic order parameter $Q$ experimentally, 
in the same material, by extending their previous study [9].
The dynamic phase transitions 
has been studied from the observed variation of $Q$.
However, the detailed study of the dynamic phase transitions by measuring
variations of associated response functions (like the ac susceptibility,
'specific-heat', correlations, relaxations etc) have not been studied
experimentally.

\section*{Acknowledgements}

Author is grateful to Jawaharlal Nehru Centre for
Advanced Scientific Research, Bangalore for
financial support and computational facilities. Author is thankful to B. K.
Chakrabarti for critical reading of the manuscript and helpful
discussions.

\centerline {\large {\bf Figure Captions}}

\noindent Fig.1 Schematic diagram of the dynamic phase boundary in the
field amplitude ($h_0$) and temperature ($T$) plane. The dotted line is
the boundary of the discontinuous transition and the solid line represents
the boundary of continuous transition. The small circle represents the
tricritical point (TCP). Insets demonstrate the breaking of the symmetry
of the dynamic hysteresis ($m-h$) loop due to dynamic transition.

\noindent Fig.2 Monte Carlo results of the temperature ($T$) variation of
'relaxation' time ($\Gamma$) for two different values of field amplitudes
($h_0$):the bullet represents $h_0$ = 1.0 and the diamond represents 
$h_0$ = 0.5. Solid lines show the temperature ($T$) variations of dynamic
order parameter $Q$. Inset shows a typical 'relaxation' of $Q$ plotted 
against the number of cycles ($n$). The solid line is the best fit
exponential form of the data obtained from MC simulation. Here, $L$ = 100,
$\omega = 2 \pi \times 0.04$.

\noindent Fig.3 Mean field results of the temperature ($T$) variation of
'relaxation' time ($\Gamma$) for two different values of field amplitudes
($h_0$): the filled triangle represents $h_0$ = 0.4 and filled square
represents $h_0$ = 0.3. Solid lines represent the temperature variation
of the dynamic order parameter $Q$. Inset shows a typical 'relaxation'of
$Q$ plotted against the number of cycles $n$. The solid line is the best
fit exponential for of the data obtained from the solution of equation 
\ref{mf}. Here, $\omega = 2 \pi \times 0.02$.

\noindent Fig.4 Monte Carlo results of the temperature variations of
$C_{coop}$ for two different values of $h_0$: the filled square represents
$h_0$ = 0.8 and the filled triangle represents $h_0$ = 0.4. Solid lines
represent the temperature variations of $Q$. Inset shows the temperature
variations of $E_{coop}$ for two different values of $h_0$:(I) $h_0$ = 0.8
and (II) $h_0$ = 0.4. Here, $L$ = 100, $\omega = 2 \pi \times 0.01$.

\noindent Fig.5 Monte Carlo results of the temperature variations of
$C_{tot}$ for two different values of $h_0$: the filled square represents
$h_0$ = 0.8 and the filled triangle represents $h_0$ = 0.4. Solid lines
represent the temperature variations of $Q$. Inset shows the temperature
variations of $E_{tot}$ for two different values of $h_0$:(I) $h_0$ = 0.8
and (II) $h_0$ = 0.4. Here, $L$ = 100, $\omega = 2 \pi \times 0.01$.
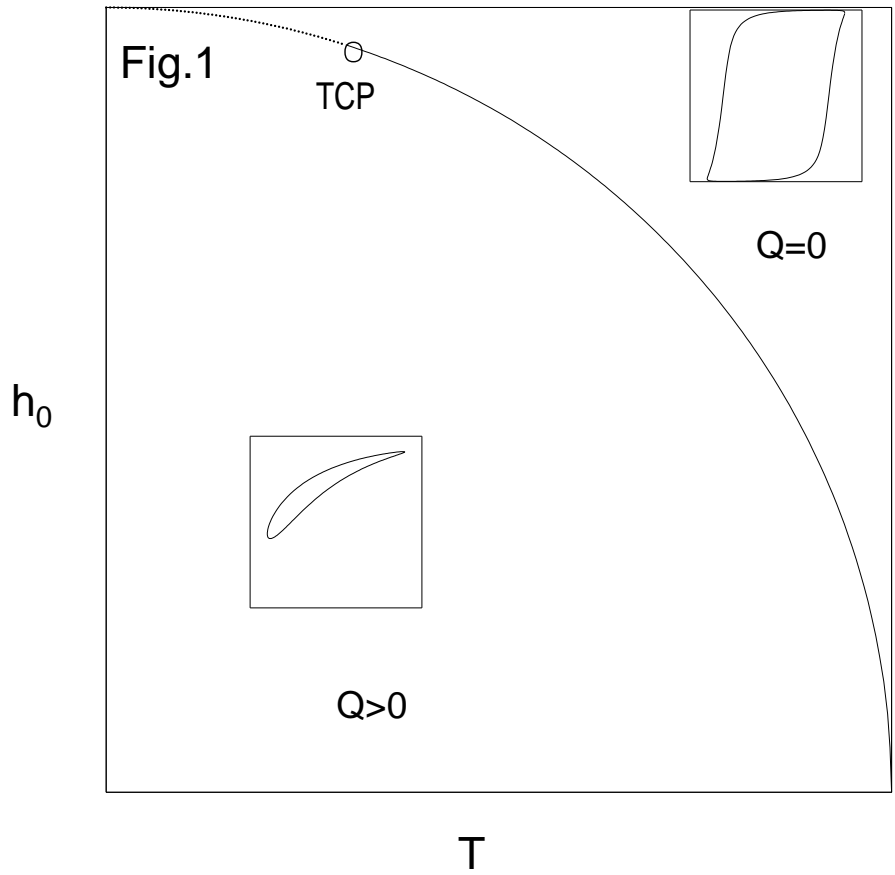
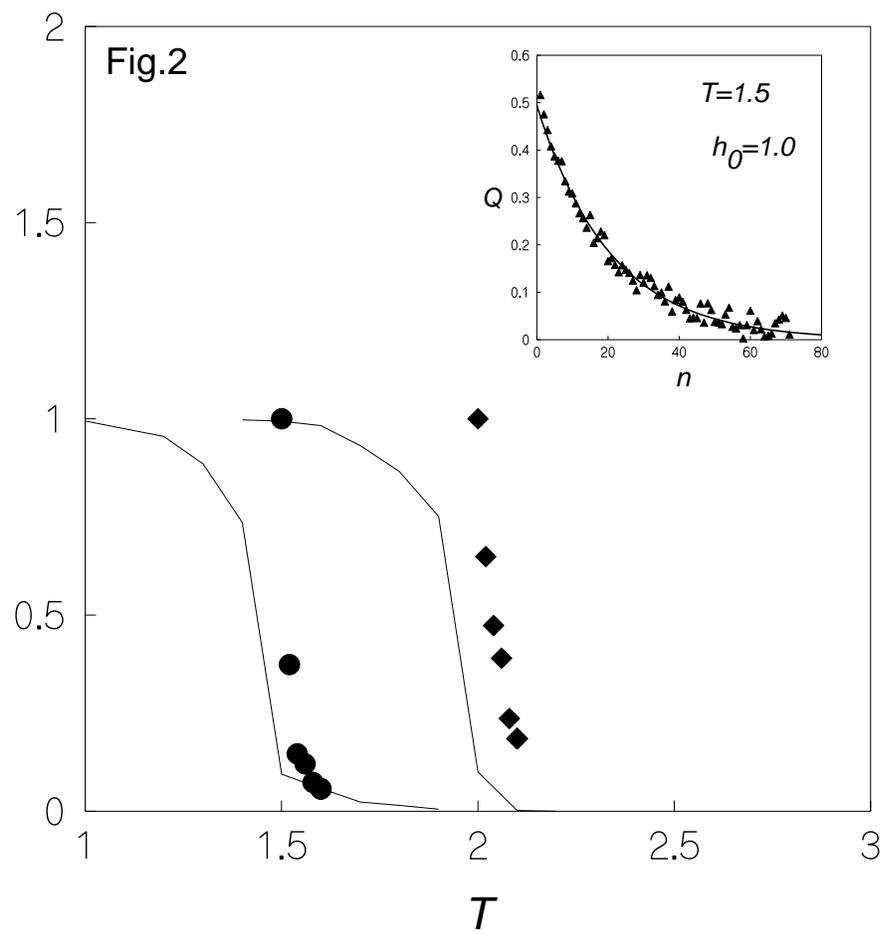
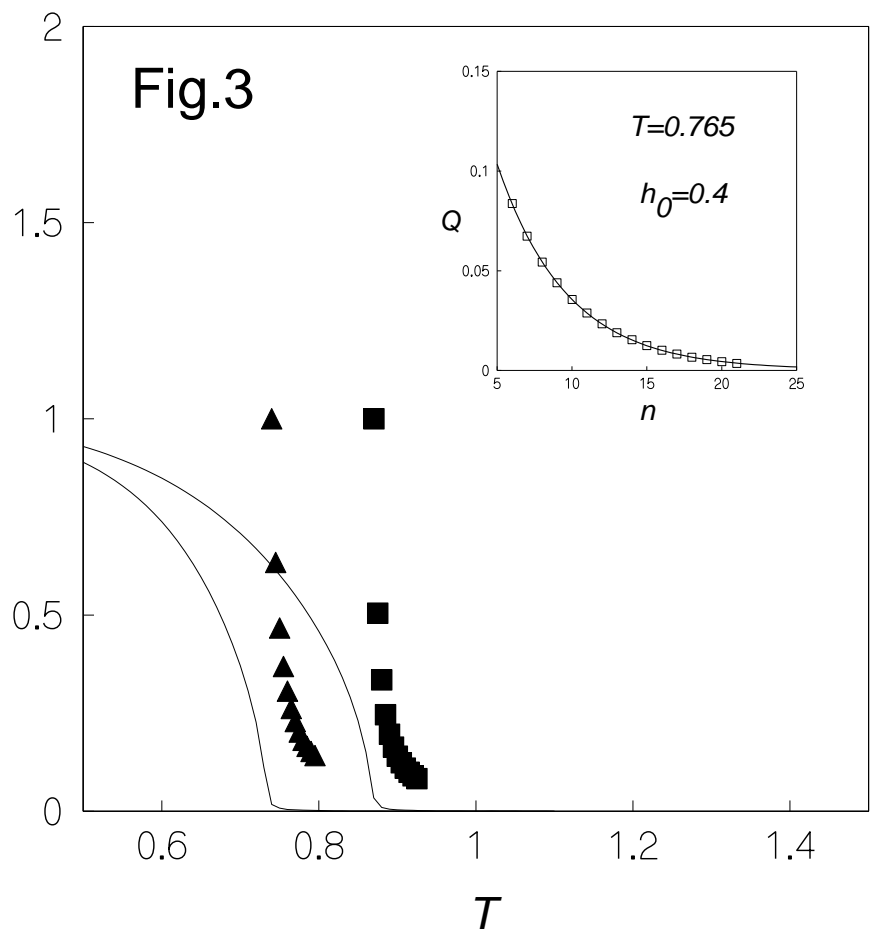
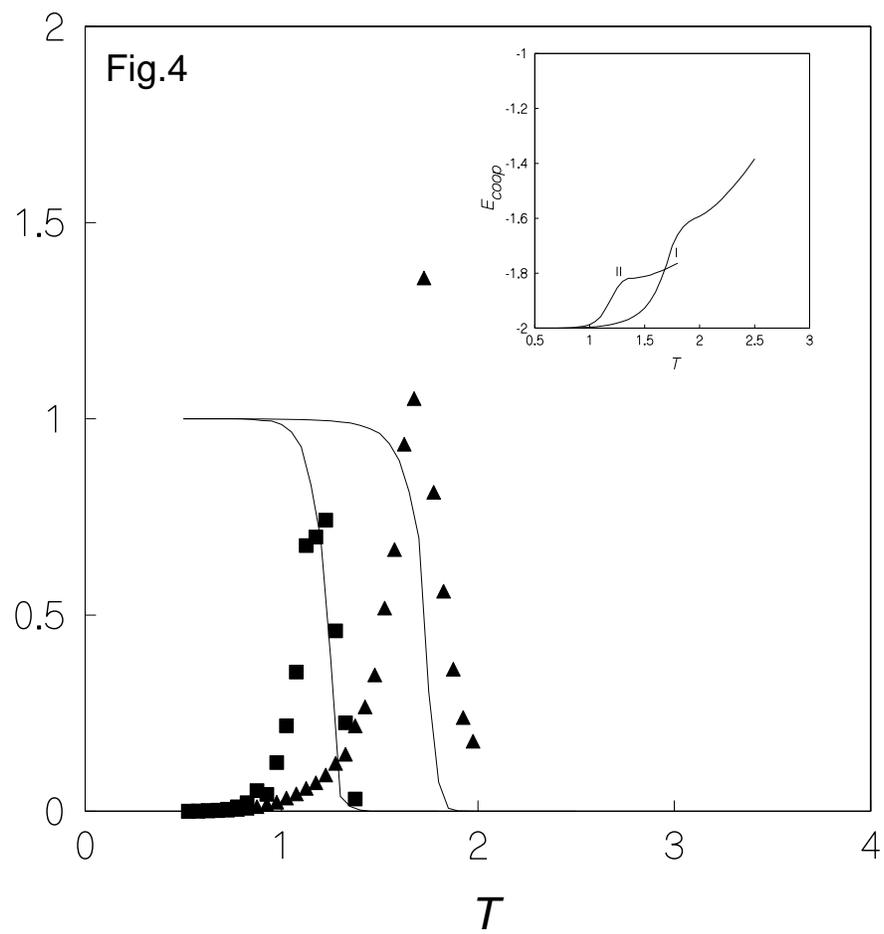
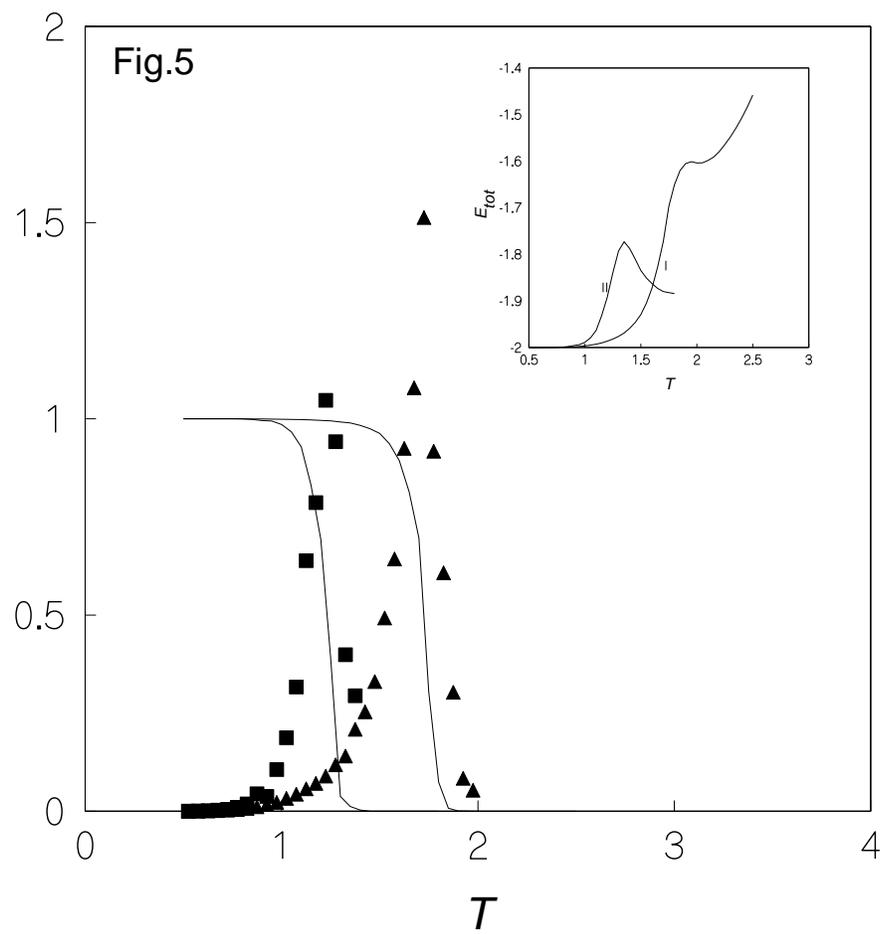
\end{document}